\documentclass[pre,twocolumn,showpacs]{revtex4-1}
\usepackage{graphicx}
\usepackage{bm}
\usepackage{amsmath}
\usepackage[percent]{overpic}
\usepackage{subfig}
\begin{document}

\title{Role of viscous friction in the reverse rotation of a disk}
\author{Pablo de Castro} 
\email{pablo@df.ufpe.br}
\author{Fernando Parisio}
\email{parisio@df.ufpe.br} 
\address{Departamento de F\'{\i}sica, Universidade Federal de Pernambuco, 50670-901,
Recife, Pernambuco, Brazil}

%\date{\today}

\begin{abstract}
The mechanical response of a circularly-driven disk in a dissipative medium is considered.
We focus on the role played by viscous friction in the spinning motion of the disk, especially 
on the effect called reverse rotation, where the intrinsic and orbital rotations are antiparallel. 
Contrary to what happens in the frictionless case,
where steady reverse rotations are possible, we find that this dynamical behavior may exist only as 
a transient when dissipation is considered. Whether or not reverse rotations in fact occur depend on the initial conditions 
and on two parameters, one related to dragging, inertia, and driving, the other associated with the geometric configuration
of the system. The critical value of this geometric parameter (separating the regions where reverse rotation is possible
from those where it is forbidden) as a function of viscosity is well adjusted by a q-exponential function.
\end{abstract}
\pacs{45.20.dc, 45.40.Bb, 81.40.Pq}
\maketitle

\section{Introduction}
Classical mechanics of simple, low dimensional and integrable systems can be surprisingly rich, provided that 
non-linearity is present. These systems, although less complex than the chaotic ones, hardly 
allow for full analytical treatments and may present behaviors like reverse rotations, excitability, transients 
and hysteresis, thus, constituting an important resource in basic physics. Not less importantly, for obvious reasons,
all sorts of machinery in industries work in non-chaotic regimes, however, displaying a variety of 
complex dynamical effects arising from non-linearities.             

One class of problems that has been often considered in the literature is that of rigid body dynamics on 
flat surfaces. Farkas et al have explored the subtle connection between translational and
spinning motions of a free disk (of radius $R$) moving on a surface with Coulombian friction \cite{farkas}.
They showed, e. g., that the terminal translational ($v$) and rotational ($\Omega$) velocities,
vanish simultaneously, and that the ratio  $\epsilon=v/R\Omega$ always tends to $\approx 0.653$ in the imminence 
of stopping, no matter its initial value (see also \cite{sliding1,sliding2}).
Complementarily, the motion of driven disks sliding on frictionless surfaces has been shown to be quite non-trivial 
regarding a dynamical behavior called reverse rotation \cite{reverse1}.

The position of a rigid body in two dimensions is completely characterized 
by the location of its center of mass (c.m.) and by the angle between some reference line marked on the body and an arbitrary 
coordinate axis. A reverse rotation develops when the c. m. follows a bounded trajectory in, say, the clockwise 
direction and, at the same time, the intrinsic angular degree of freedom evolves counterclockwise, or vice-versa.
Thus, a reverse rotation is characterized by antiparallel orbital and spin rotations. 
Famous examples of such a phenomenon are the reverse, or retrograde, rotations of Venus \cite{venus,correia1,correia2} and Uranus \cite{venus}.
Behaviors belonging to the same class can be found in the dynamics of rolling cylinders immersed in viscous fluids 
\cite{cylinder,cylinder2,cylinder3} and in the chaotic response of a damped pendulum parametrically excited \cite{chaos}.

From a more applied point of view, reverse rotations may appear, being potentially deleterious, in bearings of journal machinery \cite{decamilo}.  They also seem to be relevant in the problem of biological tissue production, where 
a commonly 
used method to generate tissue is the rotating vessel bioreactor. It consists of a cylindrical container rotating about its longitudinal axis with constant angular speed. A porous disk is seeded with cells to be cultured, 
placed within the bioreactor which, in turn, is filled with a nutrient-rich medium. The rotating fluid keeps the growing tissue construct suspended against gravity and leads to intricate dynamical regimes.
Various studies on the orbits described by the disk exist \cite{cummings, cummings2}. Although this system is not identical to the one we study here, it is clear that a better understanding of the regimes of intrinsic rotation, in the spirit of the present work, is needed. For example, the existence of a transient between two distinct spinning regimes, like the one we find here, would produce ``topological'' defects in the tissue. 

In this work we consider both, driving and friction, simultaneously. 
Our main goal is to understand how the presence of viscous friction affects the regimes of
reverse rotation that have been shown to exist for circularly-driven disks in the non-dissipative case \cite{reverse1}.
Hereafter we refer to the regime where both angular momenta are parallel as normal or prograde.

\section{System and equation of motion}
\begin{figure}
\includegraphics[width=6cm,angle=0]{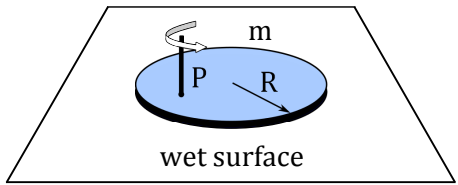}
\caption{(color online) Pictorial representation of the proposed model system}
\label{figure1}
\end{figure}
\begin{figure}
\includegraphics[width=4cm,angle=0]{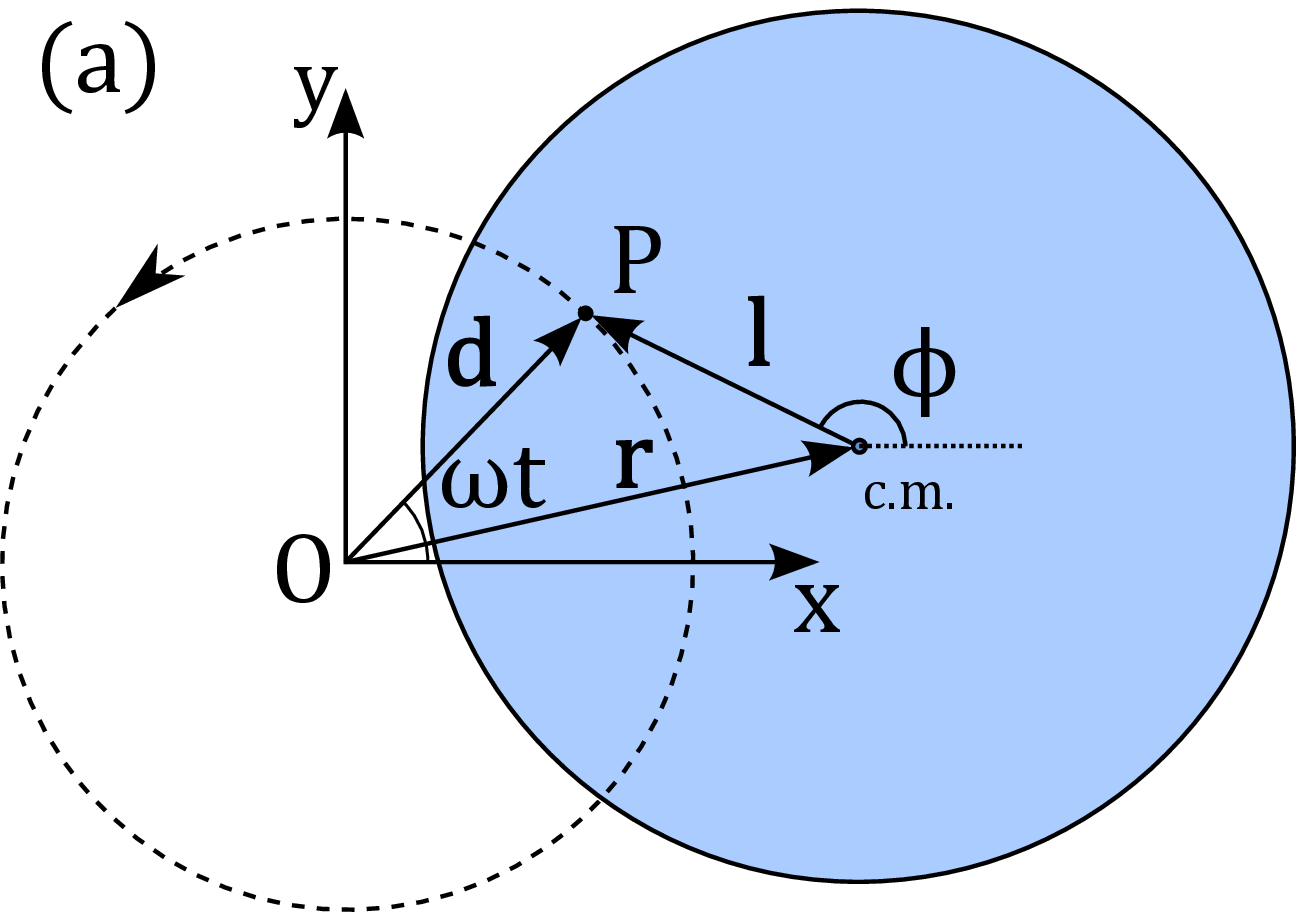}
\includegraphics[width=3.2cm,angle=0]{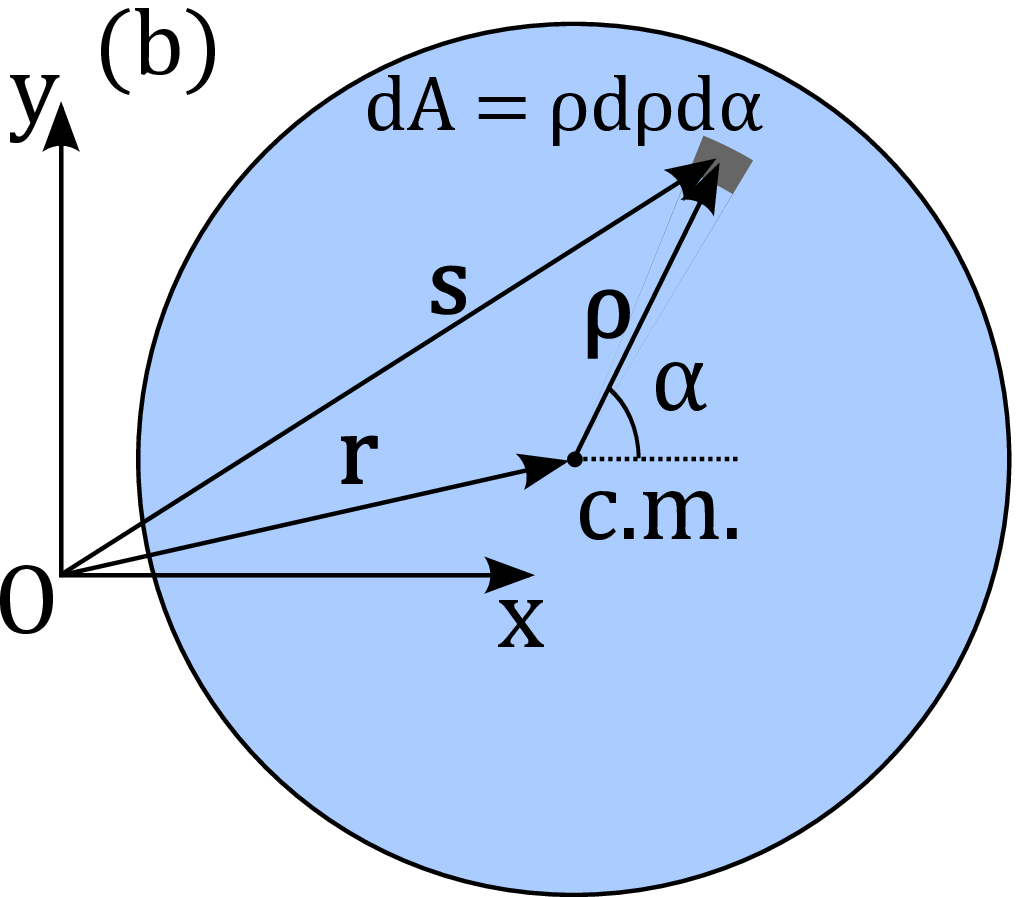}
\caption{(color online) Schematic upper view of the system where the relevant geometric 
quantities are depicted (a) and integration variables for the calculation of the viscous force and torque (b).}
\label{figure2}
\end{figure}

Our model system is depicted in Fig. \ref{figure1}. It
consists of a uniform disk of mass $m$ and radius $R$, initially resting on a
horizontal surface. The system is submitted to
an external horizontal force, provided by a driving
mechanism, through a thin rod attached to a fixed point ($P$) on
the disk, around which the whole body can rotate freely. The
driving apparatus takes the disk from rest and makes the point $P$
follow a uniform circular trajectory of radius $d$ around a
fixed origin ($O$) with angular frequency $\omega$ [see Fig.
\ref{figure2} (a)]. For definiteness we assume the rotation to be
counterclockwise and, without loss of generality, we use a
coordinate system for which the point $P$ lies in the positive
$x$-axis at $t=0$. 

For later times we denote the position vector
of $P$ by ${\bf d}$ and the vector locating the c. m.
by ${\bf r}$. Since the disk is assumed to be perfectly
rigid, $P$ is always a distance $l$ apart from c.m. The relative
position of these two points is given by the vector ${\bf l}$, as
shown in Fig. \ref{figure2} (a). Finally, the angle between the $x$-axis
and the line connecting c.m. and $P$ is denoted by $\phi$. The
variables ${\bf r}$ and $\phi$ completely specify the
position of the disk, while $\omega$ reflects the strength of driving. 

In a previous work the scenario of negligible friction 
was assumed \cite{reverse1}. 
Here we remove this restriction
by considering that a thin layer of fluid exists between the disk and the
horizontal surface, giving rise to wet friction. This
amounts to viscous forces and torques that are proportional to the local 
relative velocity between
the surfaces.
\subsection{Viscous friction}
The overall viscous force acting on the disk is given by the surface integral
\begin{equation}
{\bf F}=-\frac{b}{\pi R^2}\int_{disk} \dot{\bf s} \;\rho {\rm d}\alpha{\rm d}\rho \;,
\end{equation}
where $b$ is the drag constant and ${\bf s}$ is the vector that locates the element of area $\rho {\rm d}\alpha{\rm d}\rho$ [see fig \ref{figure2}(b)].
Taking into account the constraints ${\bf r}+{\bf l}={\bf d}$ and ${\bf r}+\boldsymbol\rho={\bf s}$,
the result reduces to that of a rectilinear, irrotational, motion of the disk: ${\bf F}=-b \dot{\bf r}$. 
In addition, the forces on each element of area give rise to a torque:
\begin{equation}
{\bf T}=-\frac{b}{\pi R^2}\int_{disk} {\bf s} \times \dot{\bf s} \;\rho {\rm d}\alpha{\rm d}\rho= - b {\bf r} \times \dot{\bf r}-\frac{bR^2}{2} \dot{\phi}\hat{z} \;,
\end{equation}
where $\phi$ is the angle between the vector ${\bf l}$ and the x-axis, and $\hat{z}$ is the unit vector perpendicular to the plane of motion. Plugging these expressions in 
Newton's second law and eliminating the c. m. degrees of freedom we get
\begin{eqnarray}
\nonumber
I_P\ddot{\phi}+b\left( l^2+\frac{R^2}{2}\right) \dot{\phi}
-mdl\omega^2\sin(\phi-\omega t)\\-bdl\omega\cos(\phi-\omega t)=0\;,
\end{eqnarray}
where $I_P$ is the inertia moment relative to the pivotation point $P$. By writing 
\begin{equation}
\label{changevar}
\theta=\phi-\omega t+ \arctan\left(\frac{b}{m\omega} \right)+\pi\;,
\end{equation}
the equation of motion considerably simplifies to
\begin{equation}
\label{eqII}
{\cal S}\frac{\ddot{\theta}}{\omega^2}+\frac{\dot{\theta}}{\omega} +\frac{\sqrt{{\cal S}^2+1}}{H}\sin \theta+ 1=0\;,
\end{equation}
where the parameter 
\begin{equation}
\label{H}
H=\frac{L^2+1/2}{DL}\;,
\end{equation} 
with $D=d/R \in [0,\infty)$ and $L=l/R \in [0, 1]$, contains all the relevant information on the scale-free geometry of the system, and ${\cal S}=m \omega/b$ gives the relative strength 
of inertia and driving versus viscous forces on the disk. Although in our calculations we will use Eq. (\ref{eqII}), it is possible to obtain a formally simpler equation by using the dimensionless time $\tau=({\cal S}{\cal A}/\omega)t$, with
${\cal A}= H^{1/2}\left( {\cal S}\sqrt{{\cal S}^2+1}\right)^{-1/2}$. The resulting two-parameter differential equation reads
\begin{equation}
\label{eqIII}
\frac{{\rm d}^2\theta}{{\rm d}\tau^2}+{\cal A}\frac{{\rm d}\theta}{{\rm d}\tau}+\sin \theta+{\cal S}{\cal A}^2=0\;.
\end{equation}
The cost is that ${\cal A}$ is an involved mixture of geometry, inertia, driving and viscosity.
From Eq. (\ref{eqIII}) we see that our problem can be mapped into the dynamics of a pendulum immersed in a fluid and subjected to a constant torque \cite{pendulum}. It is interesting to note that a number of quite distinct physical systems are, in some regimes, described by the very equation (\ref{eqIII}). Examples are the dynamics of the phase difference between the collective wave functions through a Josephson junction \cite{baker,strogatz}, the excitable behavior of microparticles under the action of an optical torque wrench \cite{wrench}, and alternate currents in electrical devices \cite{tricomi,leonov}. It is, however, important to note two points. First, the physical quantity we are interested in, which defines reverse or normal rotations, is $\phi$ and not $\theta$. Second, to completely characterize the problem, we must provide physically valid initial conditions. In the present case it is natural to assume that, at first, the disk is resting on the horizontal surface, and at a certain instant, say $t=0$, the driving apparatus is turned on. If the driving mechanism is robust enough, we can assume that this initial dynamics is impulsive, that is, the pivotation point is taken from rest to the final constant angular velocity in a time interval much shorter than any other time scale in the problem. Under these conditions it has been shown in \cite{reverse1} that, given the initial angle of the static disk, the angular velocity it acquires immediately after the driving apparatus is switched on is
\begin{equation} 
\label{ic}
\dot{\phi}_0=\frac{\omega}{ H}\cos\phi_0\;.
\end{equation}
Replacing this relation in the equation of motion we also find that $\ddot{\phi}_0=(\omega^2/ H)\sin\phi_0$. In the original derivation the friction was not taken into account. This, however, does not affect the above result due to
the hypothesis of impulsivity. 

To illustrate the restrictions imposed by the previous relation we remark that the system studied 
in \cite{pendulum} presents the interesting behavior of excitability, namely,
the existence of a dynamical regime with sharp spikes in $\theta(t)$. 
The authors observe this phenomenon for a condition that, in our notation, reads
$\omega > (b/m) \sqrt{H^2-1}$ with initial condition $\dot{\theta}_0=0$, and arbitrary $\theta_0$. Replacing this condition in
(\ref{ic}) we get $H=\cos\phi_0$, which implies $H \le 1$, leading to a negative argument in the square root. Therefore, our system, 
does not present the excitability observed in \cite{pendulum} due to the initial conditions we are concerned with.

\subsection{Limit cases}
We start this subsection with a summary of the main conclusions obtained in the frictionless ($b=0$) case
\cite{reverse1}. It was found that a regime of perennial reverse rotations is possible for an interval
of initial angles $\{\pi-\phi_B, \pi+\phi_B\}$ centered at $\pi$ (the subscript $B$ stands for ``boundary"), provided that the geometrical constraint 
\begin{equation}
\label{constraint}
H<0.793
\end{equation}
is satisfied. In fact, it can be derived from Eqs. (5) and (11) of \cite{reverse1} that, fixed the angle $\phi_B$, the critical value of $H$ below which reverse rotations occur can be determined by the solution of the transcendental equation
\begin{equation}
\label{transc}
K\left(\frac{2\sqrt{H}}{\sqrt{H^2+2H+\cos^2\phi_B}}\right)=\frac{\pi}{2H}\sqrt{H^2+2H+\cos^2\phi_B}\;,
\end{equation}
where $K$ denotes the complete elliptic function of the first kind. For $\phi_B=0$, the only angle leading to reverse dynamics is $\phi_0=\pi$. In this case the above equation reduces to $K(2\sqrt{H}/(H+1))=\pi(H+1)/2H$, whose non-trivial solution is $H_c=0.793$. For all other situations, where $\phi_B \ne 0$, $H_c$ assumes smaller values. For $H>0.793$ no initial condition may develop a reverse rotation.
Note that this necessary and sufficient condition does not depend on the 
driving frequency $\omega$, on the mass $m$ and on the absolute values of the lengths $d$, $l$, and $R$.
Initial conditions that do not satisfy the previous requirements, either lead to a permanent regime of normal rotations or, 
in the boundary between the two regimes, to an oscillatory motion with a vanishing average of $\phi$ as time goes by (see figure 3 of \cite{reverse1}).

In the opposite limit $b \rightarrow \infty$, Eq. (\ref{eqII}) becomes
\begin{equation}
\label{blarge}
\dot{\theta}+\frac{\omega}{ H} \sin \theta+\omega=0\;,
\end{equation}
whose fixed points are given by $\sin \theta^*=-H$ and $\cos \theta^*=\pm \sqrt{1-H^2}$. Let us summarize the three qualitatively distinct solutions of this equation \cite{strogatz}. (i) If $H<1$ we have two fixed points in the interval $[0,2\pi)$, the stable one having $\cos \theta^*>0$, and no oscillation. (ii) For $H=1$ we have a saddle-node bifurcation with no oscillation for any finite time. (iii) If $H>1$ there are no fixed points and the system oscillates with a period given by 
\begin{equation}
T=\frac{2\pi}{\omega\sqrt{1-H^{-2}}}\ge \frac{2\pi}{\omega}\;.
\end{equation}
The important point is that, by inspecting Eq. (\ref{changevar}), we note that, in any case, $\phi(t)$ is an increasing function of time, on average, thus, presenting normal rotations only. 

Therefore, we conclude that while in the frictionless case reverse rotations may occur, they are forbidden in the high viscosity limit. The question arises what happens in the midway?
\section{Arbitrary viscosity}
\begin{figure}
\includegraphics[width=9.5cm,angle=0]{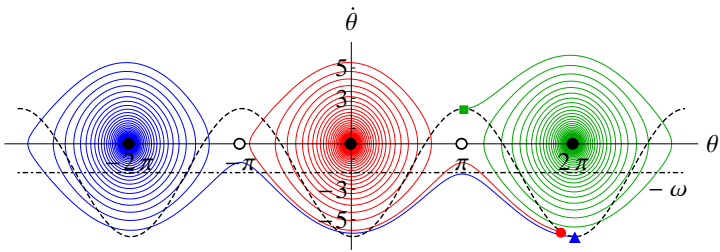}
\caption{(color online) Phase-space trajectories for three different initial conditions. More details in the text.}
\label{fig3}
\end{figure}

Although a closed analytical solution for our problem with arbitrary viscosity is not available, we can establish the stationary points and their stability properties before going into numerical solutions. By setting $\ddot{\theta}=0$ and $\dot{\theta}=0$ in (\ref{eqII}) we obtain stable fixed points satisfying
\begin{equation}
\theta^*=-\arcsin\left(  \frac{H}{\sqrt{{\cal S}+1}}\right)\pm2n\pi,
\end{equation}
with $n=0,1,2,\dots$ The unstable stationary points are given by $\pi-\theta^*$.
Our numerical investigation for intermediate values of the parameter $b/m$ begins with some typical trajectories in the plane $\theta-\dot{\theta}$. We employed the auxiliary variables directly because only in terms of them the equation of motion does not contain time explicitly. In Fig. \ref{fig3} we show a phase-space diagram for three initial conditions $\theta_0\approx 2 \pi$ ($\phi_0=\pi$) for the triangle, $\theta_0\approx \pi$ ($\phi_0=0$) for the square, and $\theta_0=2\pi-\pi/8$ ($\phi_0=\pi-\pi/8$) for the circle, with the corresponding initial velocities given by relation (\ref{ic}), represented by the dashed line in the diagram. The other parameters are as follows, $b/m=0.09$, $\omega=1.9$, and $H=0.45$. This leads to $\sin \theta^*=-0.021$, $\theta^*\approx -1,2^o$. The horizontal dashed-dotted line represents the negative of the driving angular frequency $\omega$. Recall that, to have a reverse rotation ($\langle \dot{\phi}(t) \rangle<0$ over a time scale of $2\pi/\omega$), we must get $\langle \dot{\theta}(t) \rangle<-\omega$ over a time scale of $2\pi/\omega$. We note that the first initial condition immediately leads to normal rotation, for the second condition a reverse behavior develops during a single cycle, while the thrid initial condition presents two cycles of reverse rotation before spiraling to the leftmost stable fixed point.

\begin{figure}
\includegraphics[width=7cm,angle=0]{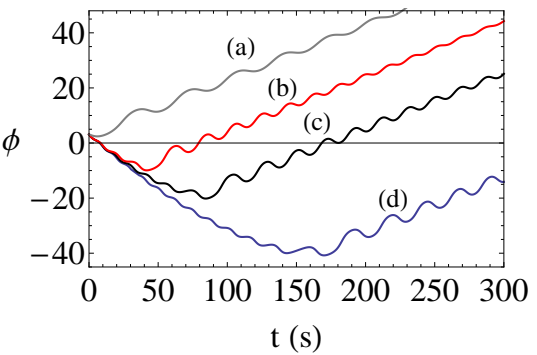}
\caption{(color online) The angle $\phi$ as a function of $t$ for $\omega=0.2$ rad s$^{-1}$, and $\phi_0=\pi$. Curve (a) presents a prograde dynamics, as expected, since $H=1.0>H_c$. For $H=0.3$ and $b/m=0.014$ s$^{-1}$ (b), $b/m=0.008$ s$^{-1}$ (c), and  $b/m=0.004$ s$^{-1}$ (d), reverse rotations are observed.}
\label{fig4}
\end{figure}

Going to the physical variable $\phi$, the first important thing to note is as follows. For every initial condition that started to develop a reverse rotation, after some time, the spin invariably flips to a regime of prograde rotation for any $b\ne 0$. This can be understood by noting that the oscillations in $\phi(t)$ eventually fade out, with $\ddot{\phi}\rightarrow 0$ for sufficiently long times, an effect observed in all investigated configurations. When this regime is reached the equation of motion becomes identical to Eq. (\ref{blarge}), for which reverse rotations have been shown to be absent. Therefore, at some moment reverse dynamics is replaced by normal rotation, the smaller the drag the longer the flip time, $t_f$, which is determined by the global minimum of $\phi(t)$. In fig. \ref{fig4} we show $\phi$ as a function of $t$ for $\omega =0.2$ rad s$^{-1}$, and initial condition $\phi_0=\pi$ (the most favorable to reverse rotations). The first curve (a) occurs for $H=1.0$ (no reverse motion) and $b/m=0.008$ s$^{-1}$. The three other curves refer to $H=0.3$ with distinct values of the drag parameter: $b/m=0.014$ s$^{-1}$ (b), $b/m=0.008$ s$^{-1}$ (c), and  $b/m=0.004$ s$^{-1}$ (d). 
We, therefore, conclude that no steady reverse rotation is allowed for any finite value of $b/m$. This regime, however, can exist as a transient that lasts longer for smaller values of viscosity. 

\begin{figure}[!ht] 
    \subfloat{%
      \begin{overpic}[width=0.43\textwidth]{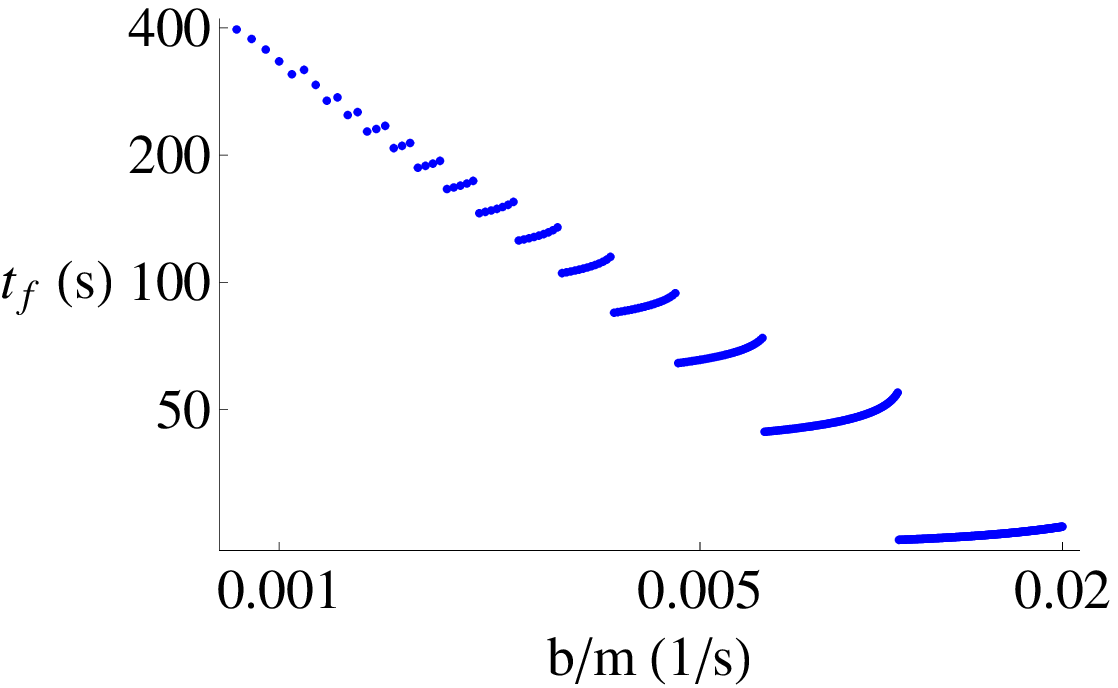}
        \put(52,32){\includegraphics[width=45\unitlength]{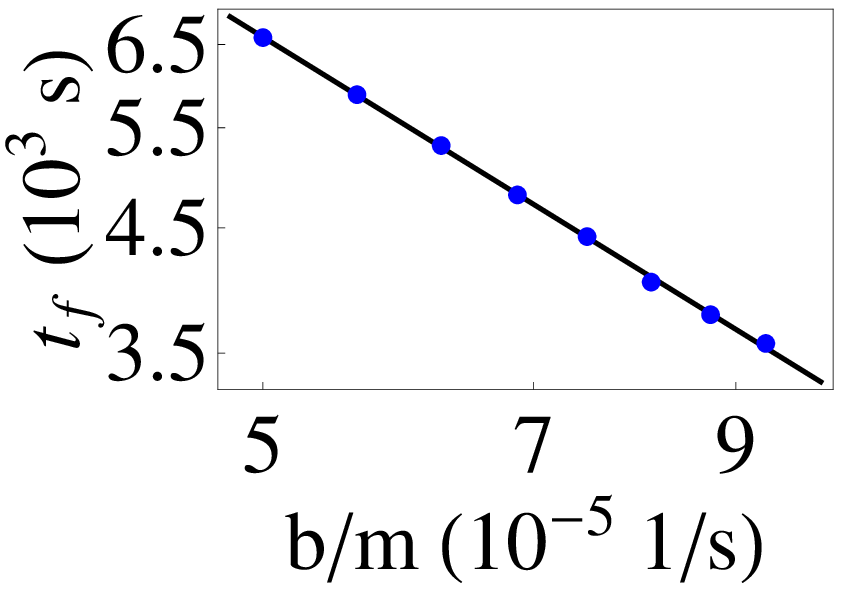}}  
      \end{overpic}
    }
    \caption{(color online) Log-log plot of the flip time $t_f$, as a function $b/m$, showing a power-law behavior in the regime of low viscosity. The discontinuities are due to the passage of the global minimum through consecutive local minima. The inset depicts a region of much weaker viscosity.}
\label{figure5}
\end{figure}

The life-time of reverse rotations must be null in the limit of high viscosity and has to diverge in the frictionless regime. Dimensional analysis leads us to infer that, if this divergence is described by a power law, then we should have $t_f \sim \omega^{\gamma-1}(b/m)^{-\gamma}$. In Fig. \ref{figure5} we record $t_f$ as a function of $b/m$ in a log-log plot. The discontinuities happen when the global minimum jumps between neighbor local minima.  Despite these jumps, a linear backbone is noticeable in a broad range of viscosity values, and a power-law divergence in the low viscosity limit is clear. The obtained relation is
\begin{equation}
\label{powerlaw}
t_f\sim \left( \frac{b}{m}\right)^{-1}
\end{equation}
which we found to be asymptotically independent of $\omega$ \cite{comment2}. The variation of the drag parameter has a much less dramatic effect on the regime of normal rotations, the only difference being the rate at which the amplitude of the oscillations in $\phi(t)$ goes to zero. 

\begin{figure}[!ht] 
    \subfloat{%
      \begin{overpic}[width=0.43\textwidth]{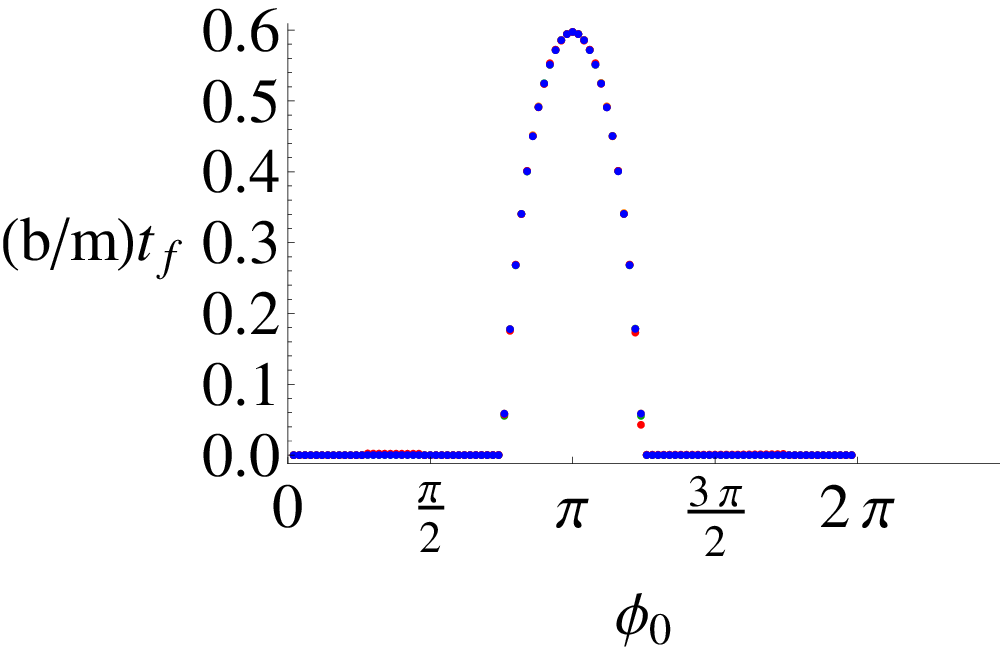}
        \put(66,26){\includegraphics[width=33\unitlength]{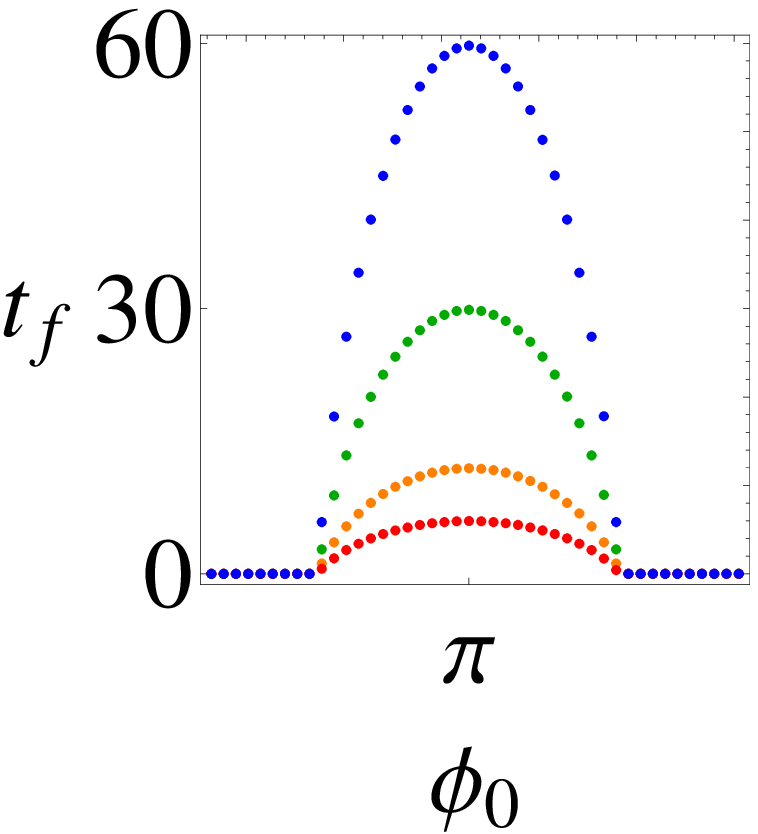}}  
      \end{overpic}
    }
    \caption{(color online) Dimensionless flip time $(b/m)t_f$ as a function of the initial condition $\phi_0$ for four different values of $b/m$.
The plots completely overlap. The inset shows the uncollapsed curves of $t_f$ against $\phi_0$. Further details in the text. }
\label{figure6}
\end{figure}

In spite of this, we verified that the initial conditions that lead to reverse rotations in the case of vanishing viscosity are the same that give rise to the reverse transient, the interval $\{\pi-\phi_B,\pi+\phi_B\}$ being insensitive to the value of $b$. This is expected due to the impulsive nature of the initial energy input. In addition, due to relation (\ref{powerlaw}) we suspected that the dependence of $t_f$ with $\phi_0$ might be universal with respect to the dimensionless time $(b/m)t_f$. This is indeed the case, as figure \ref{figure6} shows for different values of $b/m$. In the inset we plot the uncollapsed curves of $t_f$ alone against $\phi_0$.  We used a high angular velocity, $\omega=200$ rad s$^{-1}$, in order to suppress oscillations and make the plot clearer.  For lower values of $\omega$ the results are qualitatively the same and fig. \ref{figure6}
would represent the envelope of the actual plots. The other parameters employed are $H=0.3$, and $b/m=0.1$ s$^{-1}$ (a), $b/m=0.2$ s$^{-1}$ (b), $b/m=0.5$ s$^{-1}$ (c), $b/m=1.0$ s$^{-1}$ (d).  The invariant value of $\phi_B$ is approximately
$3\pi /5$

We now turn our attention to the geometric parameter defined in (\ref{H}). In the absence of friction, we found that reverse
rotations are possible only if $H<0.793=H_c(0)$, according to (\ref{constraint}). Although perennial reverse rotations are not present for $b\ne 0$, one may ask which values of $H$ allow for transient reverse behavior. In Fig. \ref{figure7} we display the maximum value of $H$ below which reverse dynamics can occur as a function of $b/m \omega$, with $\phi_0=\pi$. Interestingly enough, the numerical data are very well described by a q-exponential function :  
\begin{figure}
\includegraphics[width=7cm,angle=0]{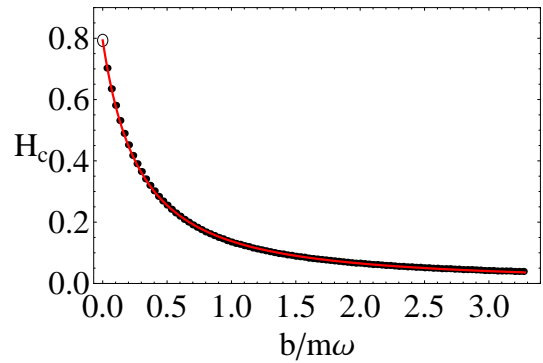}
\caption{(color online) Critical geometric parameter $H_c$ versus the dimensionless variable ${\cal S}^{-1}=b/m\omega$. $H_c$ falls off following a q-exponential function with $q=1.7$ for $\phi_0=\pi$. The open circle represents $H_c=0.793$.}
\label{figure7}
\end{figure}
\begin{equation}
\label{qexp0}
H_c(b/m\omega)=H_c(0)\left[1-\lambda(1-q)\left(\frac{b}{m\omega}\right)\right]^{1/(1-q)}\;,
\end{equation}
with $q=1.7$ and $\lambda=3.47$. When $\phi_0$ departs from $\pi$, both, the values of $q$ and $H_c(0)$ tend to decrease. For $\phi_0=2.5$, e.g., we get $q \approx 1.5$ and $\lambda \approx 2.23$, while $H_c(0)$ considerably drops to 0.40. The values of $H_c(0)$ quickly become very restrictive to reverse motion as, for example, $\phi_0=2.0$, leading to $H_c(0)\approx 0.09$. For $L=0.5$, such a condition on $H$ would be satisfied only for $D \ge 17$, hindering in practice the occurrence of reverse rotations. We, thus, see that increasing wet friction not only decreases the life time of reverse rotations, but also reduces the region in the space of parameters $L$ and $D$ for which they are possible.
\section{Conclusions}
In this work we studied the influence of wet friction in the circularly driven motion of a disk. We found that the spinning dynamics of the disk is given by a combination (competition) between a uniform motion and a pendular motion (associated with a pendulum immersed in a viscous fluid and acted upon by a constant torque), see Eq. \ref{eqIII}. While in the frictionless case reverse rotations may exist in steady regimes, for any finite value of viscous damping, this behavior becomes a transient, thus, having a finite life time. This transient have been completely characterized: (i) Its life time follows a power law with $t_f \sim m/b$; (ii) the presence of viscosity reduces the possible geometric configurations that lead to rotations that are initially reverse due to the fall of $H_c$, according to a q-exponential; (iii) however, the interval of initial conditions $\phi_0$ that leads to retrograde behavior is insensitive to the value of $b$. In fact the whole shape of the function $(b/m)t_f(\phi_0)$ is independent of the dragging. A natural extension of the present work is to consider the analogous situation with Coulombian (dry) friction. This leads to  a more complex equation of motion involving elliptic functions of intricate arguments and requires a full numerical treatment. 

Regarding the appearance of a q-exponential (introduced in the context of statistical physics by Tsallis \cite{tsallis}) describing the behavior of a critical parameter in a situation that does not explicitly involve statistics, it might look unexpected. Although the classical foundations of non-extensive statistical mechanics may be formally understood via generalizations of the Langevin equation \cite{beck}, where viscous friction plays an essential role, one cannot easily relate our result to this kind of microscopic description. A more plausible possibility is simply attributed to the ability of q-exponentials to fit a broad class of decreasing functions, as exemplified in the appendix. Whether Eq. (\ref{qexp0})  is a pure mathematical fact, as we strongly believe, or has a deeper statistical explanation, is a matter to be investigated. The same observation is valid for the range of values we obtained for $q$, which also appears in the statistical studies of complex systems, e. g., in the distribution of urban agglomerates in Brazil and USA \cite{cities}. 
\begin{acknowledgements}
The authors thank Tiago Ara\'ujo and Victor Pedrosa for many stimulating discussions on this work.
Financial support from Conselho Nacional de Desenvolvimento Cient\'{\i}fico e Tecnol\'ogico (CNPq), Coordena\c{c}\~ao de Aperfei\c{c}oamento de Pessoal de N\'{\i}vel Superior (CAPES), and Funda\c{c}\~ao de Amparo \`a Ci\^encia e Tecnologia do Estado de Pernambuco (FACEPE) (Grant No. APQ-1415-1.05/10) is acknowledged.
\end{acknowledgements}
\appendix*

\section{q-exponentials without statistics}

In this appendix we exemplify how q-exponentials may appear, quite naturally, in problems that have no connection to statistical mechanics.

Suppose you have very little knowledge about a decreasing function $f(x)$, for instance, its value $f(0) \ne 0$ and its derivative $f'(0)$ at $x=0$. In this poor scenario the best thing one can do to extrapolate the values assumed by $f(x)$ for $x \ne 0$ is to write 
\begin{equation}
\label{taylor}
f(x) \approx f(0)+f'(0)x. 
\end{equation}
How should we proceed if we are given one more, incremental piece of information, e. g., the value assumed by $f$ at a different point $x=x_0$? 

Let us take a seemingly circular approach, namely, to estimate (also to first order) the function $F(x)=[f(x)]^Q \approx \tilde{F}(x)$, and then to write $f(x) \approx [\tilde{F}(x)]^{1/Q} \equiv \tilde{f}(x)$. First we have
\begin{equation}
\tilde{F}(x)= F(0)+F'(0)x, 
\end{equation}
with $F(0)=[f(0)]^Q$ and $F'(0)=Q[f(0)]^{Q-1}f'(0)$. That is
\begin{equation}
\tilde{F}(x)= [f(0)]^Q+Q[f(0)]^{Q-1}f'(0)x, 
\end{equation}
which amounts to
\begin{equation}
\tilde{f}(x)/f(0)=\left[1-Q\frac{|f'(0)|}{f(0)}x\right]^{1/Q}. 
\end{equation}
The above expression is exactly the definition of the q-exponential:
\begin{equation}
\label{qexp}
\tilde{f}(x)=f(0)\left[1-Q\lambda x\right]^{1/Q} = f(0) \, {\rm exp}_Q(-\lambda x), 
\end{equation}
with $\lambda=|f'(0)|/f(0)$. Note that we only assumed that $f$ is a steadly decreasing function of $x$.
These observations would be mere curiosity if it weren't the fact that by expanding the previous expression to first order in $x$ we get $\tilde{f}(x)=f(0) \, {\rm exp}_Q(-\lambda x)\approx f(0)+f'(0)x$, {\it no matter the value of} $Q$. So, we get a first order approximation at least as good as (\ref{taylor}), with the difference that there is an adjustable parameter that can be used to improve the extrapolation, if one is given any extra information on $f(x)$.

As an illustration, consider the following game. Player ${\cal A}$ picks a smooth decreasing function $f(x)$ and gives player ${\cal B}$ three pieces of information, namely, $f(0)=3.927$, $f'(0)=-0.2273$, and the extra datum $f(5)=0.3561$.
The objective of ${\cal B}$ is to generate a good approximation to the actual function $f(x)$.
If ${\cal B}$ decides to use prescription (\ref{qexp}), he gets
\begin{equation}
\label{qexp2}
\tilde{f}(x)=3.927\left(1-0.0579\, Q x\right)^{1/Q}.
\end{equation}
Now ${\cal B}$ has to choose a value of $Q$. He makes a very simple decision, adjusting $Q$ so that $\tilde{f}(5)=f(5)$, which leads to $Q= -1.162$.
After that, ${\cal A}$ reveals the true function $f$: 
\begin{equation}
\label{atan}
f(x)= \frac{\arctan (1-x)+\pi}{\sqrt{1+5x}},
\end{equation}
\begin{figure}
\includegraphics[width=6.5cm,angle=0]{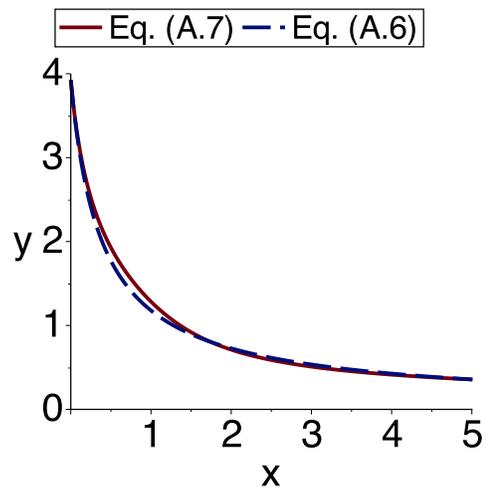}
\caption{(color online) Plots of $y=f(x)$ as given by (\ref{atan}) and of $y=\tilde{f}(x)$ as given by (\ref{qexp2}) with $Q=-1.162$. }
\label{figure8}
\end{figure}
which is indeed compatible with the information provided to ${\cal B}$: $f(0)=5\pi/4$, $f'(0)=-1/2-25\pi/8$, and $f(5)=[ \arctan (-4)+\pi]/\sqrt{26}$.
The plots of $f(x)$ and $\tilde{f}(x)$ are shown in figure \ref{figure8}, which corresponds to a quite reasonable result for such a rough procedure.
The reader is suggested to choose her (his) own function and proceed analogously. The results are poor only if $f$ decreases faster than $\exp(-\alpha x)$ or if the function has the shape of a bell (because in this case one should use a q-Gaussian).
Although our reasoning does not formally prove that q-exponentials are naturally suited to fit decreasing functions in scenarios of scarce data, it does provide evidence for this belief. In particular, based on this peculiarity, one cannot state that the actual functional form of the critical parameter $H_c$ {\it is} that of a q-exponential function of ${\cal S}^{-1}$.


\begin{thebibliography}{99}

\bibitem{farkas} Z. Farkas et al, Phys. Rev. Lett. {\bf 90}, 248302 (2003).

\bibitem{sliding1} P. D. Weidman and C. P. Malhotra, Phys. Rev. Lett. {\bf 95}, 264303 (2005).

\bibitem{sliding2} P. D. Weidman and C. P. Malhotra, Physica D {\bf 233}, 1 (2007).

\bibitem{reverse1} F. Parisio, Phys. Rev. E {\bf 78}, 055601(R) (2008).

\bibitem{venus} see, for example, http://solarsystem.nasa.gov/planets.

\bibitem{correia1} A. C. M. Correia and J. Laskar, Nature {\bf 411}, 767 (2001)

\bibitem{correia2} A. C. M. Correia, J. Laskar, and O. N. de Surgy, ICARUS, {\bf 163}, 1 (2003).

\bibitem{cylinder} J. R. T. Seddon and T. Mullin, Phys. Fluids {\bf 18}, 041703 (2006).

\bibitem{cylinder2} C. Sun et al, J. Fluid Mech. {\bf 664}, 150 (2010).

\bibitem{cylinder3} A. Merlen and C. Frankiewicz, J. Fluid Mech. {\bf 685}, 461 (2011).

\bibitem{chaos} K. Yoshida and K. Sato, Int. J. Non-Linear Mech. {\bf 33}, 819 (1998).

\bibitem{decamilo} S. DeCamilo, K. Brockwell, and W. Dmochowsky, Tribol. Trans. {\bf 49}, 305 (2006).

\bibitem{cummings} S. L. Waters et al, IMA J. Math. Med. Biol. {\bf 23}, 311 (2006) ; 
L. J. Cummings and S. L. Waters, ibid {\bf 24}, 169 (2007).

\bibitem{cummings2} L. J. Cummings et al, Biotech. and Bioeng. {\bf 104}, 1224 (2009).

\bibitem{pendulum} P. Coullet et al, Am. J. Phys. {\bf 73}, 1122 (2005).

\bibitem{baker} G. L. Baker, {\it The Pendulum: A Case Study in Physics} (Oxford University Press, Oxford, 2005).

\bibitem{strogatz} S. H. Strogatz, {\it Nonlinear Dynamics and Chaos} (Westview, Cambridge, MA, 2000).

\bibitem{tricomi} F. Tricomi, Annali della Scuola Normale Superiore de Pisa, Classe di Scienze {\bf 2}, 1 (1933).

\bibitem{leonov} G. A. Leonov,  {\it Mathematical Problems of Control Theory} (World Scientific, Singapore, 2001).

\bibitem{wrench} F. Pedaci et al, Nature Phys. {\bf 7}, 259 (2011).

\bibitem{comment2} We found a weak oscillatory dependence of $t_f$ on angular velocity for small
values of $\omega$. However, also in this situation we obtain $\gamma=1$.

\bibitem{tsallis} C. Tsallis, J. Stat. Phys. {\bf 52}, 479 (1988).

\bibitem{beck} C. Beck, Phys. Rev. Lett. {\bf 87}, 180601 (2001).

\bibitem{cities} L. C. Malacarne, R. S. Mendes, and E. K. Lenzi, Phys. Rev. E {\bf 65}, 017106 (2001).



\end{thebibliography}
\end{document}